\documentclass[9pt,twocolumn,twoside]{osajnl}
\usepackage{ol}
\usepackage{amsmath}
\usepackage{amssymb}
\usepackage{subcaption}
\newcommand\dd[0]{\mathrm{d}}
\newcommand\e[1]{\mathrm{e}^{#1}}

\hypersetup{pdfstartview={FitH},pdfpagemode={UseNone},
            colorlinks,linkcolor=blue, citecolor=blue, urlcolor=blue,
             bookmarksopen=true, pdfnewwindow=true}

\setboolean{shortarticle}{true}

\ifthenelse{\boolean{shortarticle}}{\colorlet{color2}{color2b}}{\colorlet{color2}{color2}} 

\title{Frequency comb generation in SNAP bottle resonators}

\author[1,*]{Sergey V. Suchkov}
\author[2]{Mikhail Sumetsky}
\author[1]{Andrey A. Sukhorukov}

\affil[1]{Nonlinear Physics Centre, Research School of Physics and
Engineering, The Australian National University, Canberra, ACT
2601, Australia}
\affil[2]{Aston Institute of Photonics Technology, Aston University, Birmingham B4 7ET, UK}

\affil[*]{Corresponding author: sergey.v.suchkov@anu.edu.au}

\dates{Compiled \today}
\ociscodes{(190.4223) Nonlinear wave mixing; (190.4360)
Nonlinear optics, devices; (140.3945)
Microcavities; (060.2340) Fiber optics components.}

\doi{\url{http://dx.doi.org/10.1364/ol.XX.XXXXXX}}

\begin{abstract}
We develop a theory of optical frequency comb generation in ultra-compact Surface Nanoscale Axial Photonic (SNAP) bottle microresonators, employing the nonlinear interaction of whispering gallery modes which are confined along an optical fiber with nanoscale radius variation. We predict that a SNAP microresonator with a few-$\mu$m radius can generate a frequency comb with an ultra-fine sub-GHz spectral spacing, which would require traditional ring resonators of $cm$ radius. We identify regimes of stable or quasi-periodic comb dynamics due to soliton excitation, and also show that special engineering of the SNAP radius profile can be used to compensate nonlinearity induced dispersion.
\end{abstract}

\setboolean{displaycopyright}{true}

\begin{document}

\maketitle
\thispagestyle{fancy}

\ifthenelse{\boolean{shortarticle}}{\ifthenelse{\boolean{singlecolumn}}{\abscontentformatted}{\abscontent}}{}

Photonic structures generating low repetition rate and broadband frequency combs are of great interest for a variety of applications including spectroscopy, biological and chemical sensing, and atomic clocks~\cite{Ye:2005:FemtosecondOptical}. A frequency comb can be considered as a high precision spectroscopic ruler, so that the larger is the comb bandwidth and the smaller spacing between the comb teeth is, the more accurate the ruler is.

\begin{figure}[htbp]
    \centering
    \begin{subfigure}[b]{0.8\columnwidth}
       \includegraphics[width=\linewidth]{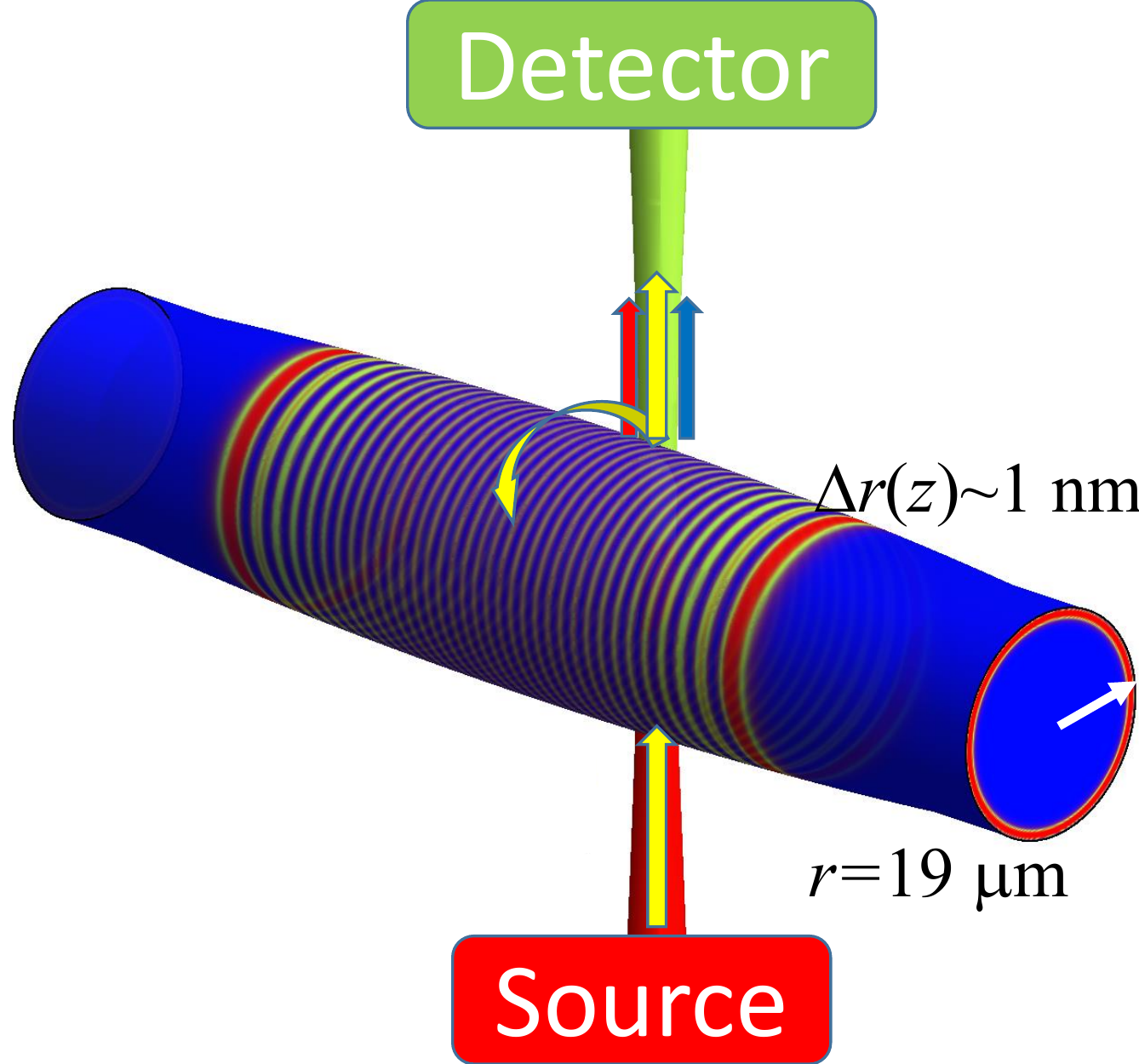}
        \caption{}
        \label{fig1a}
    \end{subfigure}
        \begin{subfigure}[b]{0.8\columnwidth}
       \includegraphics[width=\linewidth]{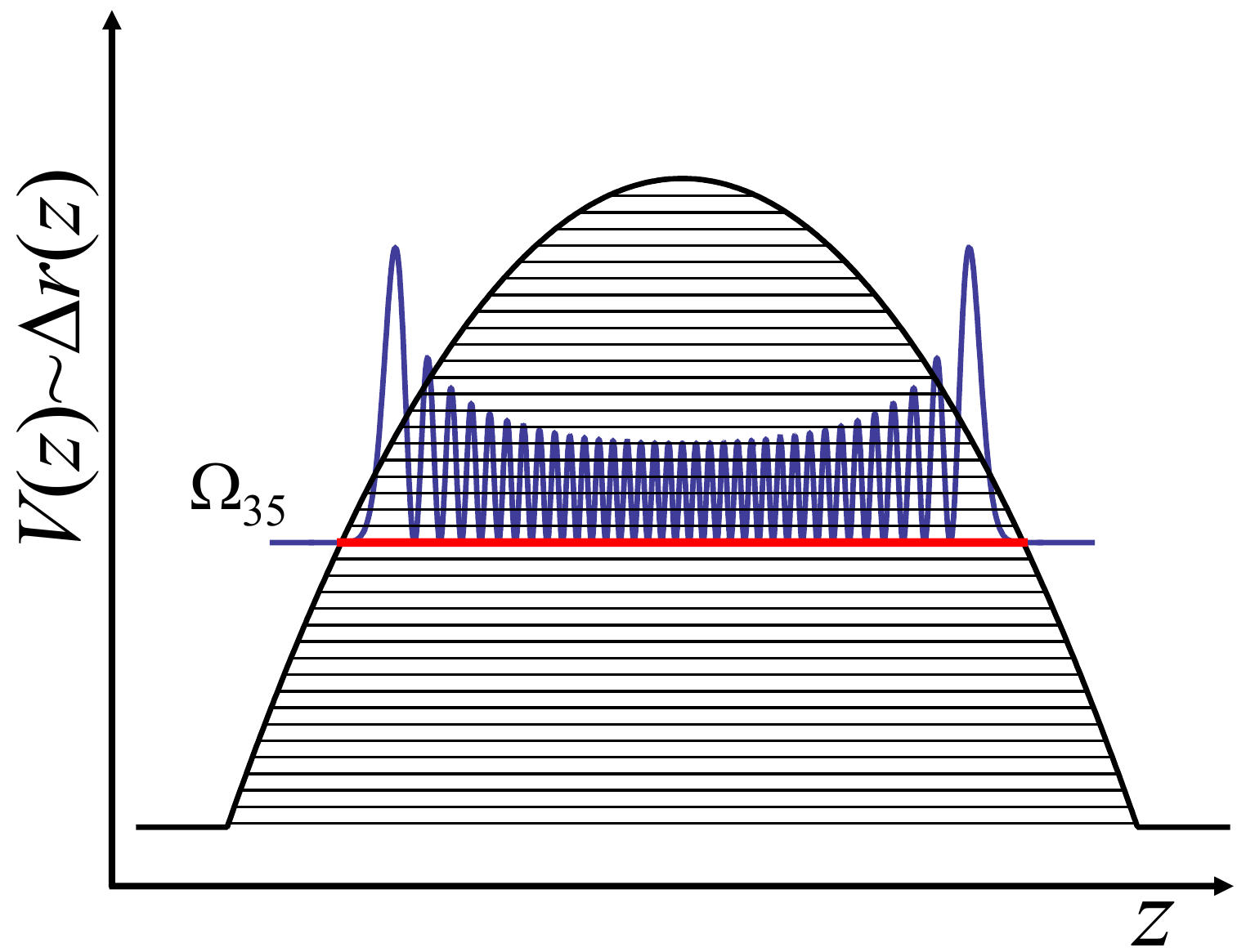}
        \caption{}
        \label{fig1b}
    \end{subfigure}
 \caption{(a) Schematic of a SNAP microresonator with a parabolic radius variation coupled to a microfiber. Characteristic intensity distributions of a WGM at the resonator surface and cross-section are shown with color shading, where blue corresponds to zero and red~-- to the maximum value.
   (b) Effective potential corresponding to the parabolic radius variation. Blue line: the axial distribution of a WGM eigenstate with eigenfrequency $\Omega_{35}$.}
    \label{Fig1}
\end{figure}

Two major approaches to obtain frequency combs are based on employing mode-locked lasers~\cite{Ye:2005:FemtosecondOptical} and nonlinear microresonators~\cite{DelHaye:2007-1214:NAT, Kippenberg:2011-555:SCI, Herr:2012-480:NPHOT}. Microresonators possess several advantages over mode-locked lasers such as smaller dimensions, lower energy consumption, and cheaper fabrication. Furthermore, the ultra-high Q-factors  $\sim 10^9$ in microresonotrs enable effectively long interaction length facilitating extremely low sub micro-Watt power thresholds for nonlinear wave mixing and comb generation.
The spacing of teeth in combs generated by microresonators is determined by their free spectral range, which is typically 2-3 orders of magnitude greater than in combs generated by mode-locked lasers. Specifically, microresonators have to be of millimeter dimensions for 10-100 GHz spacing or even larger for finer spectral combs~\cite{Kippenberg:2011-555:SCI, Johnson:2012-875:OL}, whereas the realization of fine and controlled frequency spacing in the GHz range would increase spectral precision and enable direct optical-electronic interfacing.
In this Letter we predict that Surface Nanoscale Axial Photonics (SNAP) bottle resonators~\cite{Sumetsky:2013-393:NANP} can generate low repetition rate frequency combs with GHz and sub-GHz spacing, comparable to those in mode-locked lasers, and at the same time possess very small dimensions with few-$\mu m$ cross-section.

The SNAP platform combines microscopic dimensions and ultra-low
loss with Q-factor approaching $10^8$~\cite{Sumetsky:2013-163901:PRL, Sumetsky:2013-393:NANP, Watts:2013-3249:JOSB}.
It consists of an optical fiber with specially designed nanometer-scale radius variation along the axial direction and transverse input/output microfibers coupled to the resonator, see Fig.~\ref{Fig1}(a). The microfibers are usually fabricated of bi-conical fibers with micron-scale diameter waist, or, alternatively, can be planar waveguides fabricated lithographically. An input waveguide launches whispering gallery modes (WGMs), which circulate near the surface of the SNAP resonator and in addition experience slow axial propagation with correspondingly small propagation constants, according to the concept of bottle resonators~\cite{Sumetsky:2004-8:OL, Pollinger:2009-53901:PRL, Savchenkov:2011-293:NPHOT}.

The WGMs in SNAP resonators are highly sensitivity to $nm$-scale variations of the fiber radius and the refractive index change, which can be introduced to create an arbitrary axially varying effective potential which can support multiple WGM~\cite{Sumetsky:2011-26470:OE, Watts:2013-3249:JOSB, Suchkov:2015-3806:OL}, see an example in Fig.~\ref{Fig1}(b). The potential shape can then be optimized for
efficient comb generation, by providing an equal spacing of WGM frequencies to satisfy the phase matching conditions in nonlinear four-wave mixing~\cite{Dvoyrin:2016-5547:OL}. We show that this can be achieved in a SNAP resonator with a parabolic radius variation, as illustrated in Fig.~\ref{Fig1}(b).

We analyse theoretically the frequency comb generation based on Kerr-type nonlinear four-wave mixing in SNAP resonators. While the principles of comb generation in conventional resonators are well established~\cite{Ye:2005:FemtosecondOptical, DelHaye:2007-1214:NAT, Kippenberg:2011-555:SCI}, in SNAP systems there appears an additional degree of freedom associated with axial mode dynamics. Here, we formulate nonlinear model equations which describe the mode evolution along the SNAP resonator axis ($z$), and perform extensive numerical simulations.

Both the effects of the radius variation and nonlinear wave mixing are relatively weak, and can be considered as perturbations. Then, we start our analysis by considering the properties of linear WGMs in a homogeneous SNAP resonator having a fixed radius, which electric field profiles can be expressed as
%
\begin{eqnarray}
    \label{eigenmodes}
    \vec{E}= A(z,t) \exp({-i\omega^0_m t}) \vec{e}_m(x,y),
\end{eqnarray}
%
where $A(z,t)$ are the slowly varying mode amplitudes, $\omega^0_m$ are angular eigenfrequencies of the modes $\vec{e}_m(x,y)$, which determine the field distribution in a cross section of the fiber, $m$ is an azimuthal quantum number.
Electric field profiles $\vec{e}_m(x,y)$ for the eigenmodes of an undeformed fiber in the linear regime and in polar coordinates  $(\varphi,\rho)$ can be approximated as~\cite{Sumetsky:2009-7196:OE, Sumetsky:2011-26470:OE}
\begin{equation}\label{Airy}
  \vec{e}_m=
  \begin{cases}
  \vec{E}_0 \e{i m \varphi} {\rm Ai}\left(\left[-\sqrt[3]{2m^2}{(\rho-r_0)}{r_0^{-1}} \right]-t_n \right), &  \rho\leq r_0.\\
  0, &  \rho > r_0  
  \end{cases}
\end{equation}
where $t_n$ are the roots of Airy function.
In the following we consider TM modes, such that electric field polarization is linear and vector $\vec{E}_0$ is aligned along the fiber axis $z$, and the fundamental Airy's mode with $t_0=2.3381$.

Using an established approach based on Lorenz reciprocity~\cite{Snyder:1983:OpticalWaveguide, Michaelis:2003-65601:PRE, Sukhorukov:2014-462:OL} we derive a governing equation for the mode amplitude $A$, assuming that one azimuthal mode $m$ is excited:
\begin{eqnarray}
    i\frac{\partial A}{\partial \tau}+\beta_2\frac{\partial^2 A}{\partial z^2}+\alpha|A|^2 A
    +i\Gamma A + V_{{\rm r}}(z)A\nonumber\\
    +i V_{\rm diss}(z)A =
    V_{\rm s}(z) A_{\rm s}(\tau).
\label{Sum3}
\end{eqnarray}
Here $\beta_2$ is a spatial dispersion coefficient, $N$ is a strength of nonlinear response, $\Gamma$ is intrinsic loss in the resonator, $V_{\rm r}(z)$ is a potential determined by the radius variation profile, $V_{\rm diss}(z) A$ is a term responsible for light leakage from the SNAP resonator to the microfiber, while $V_{\rm s}A_{\rm s}(\tau)$ determines light coming from the microfiber. We use dimensionless variables for the axial coordinate $z$ and time $\tau$, which are related to physical units by the scaling parameters $Z$ and $T$, respectively.
Then, we consider a WGM propagating close to the fiber surface and determine the linear parameters following Ref.~\cite{Sumetsky:2011-26470:OE}:
\begin{eqnarray}
  \beta_2=\cfrac{T}{Z^2}\cfrac{\omega_0}{2\beta^2}\ ,\quad  \Gamma = T\frac{\gamma \omega_0^2}{2\pi c},\quad
  \label{NonCef}
   V_{{\rm r}}(z)=T\omega_0\frac{\Delta r(z)}{r_0},\\
  V_{\rm diss}(z)=T\frac{\omega_0}{2\beta^2}f(z) D\ ,\quad   V_{\rm s}(z)=T\frac{\omega_0}{2\beta^2}f(z) C,
\end{eqnarray}
while the nonlinear coefficient is found as
\begin{eqnarray}
  \alpha =T\omega_0\frac{\Delta n_{nl}}{n_0} 
  {\int\limits_{0}^{r_0}\left|\vec{e}_m\right|^4\rho\, \dd \rho}  
    \times \left[{\int\limits_{0}^{r_0}|\vec{E}_0|^2 \left|\vec{e}_m\right|^2\rho\, \dd\rho}\right]^{-1},
  \label{potet}
\end{eqnarray}
where $r_0$ is the SNAP fiber radius, $\lambda_0 = 2 \pi c / \omega_0$ is the resonant wavelength,
$\gamma$ is the loss coefficient,
$n_0$ is the SNAP fiber refractive index, and $\beta=2\pi n_0/\lambda_0$ is the propagating constant. Parameters $C$ and $D$ determine light coupling `in'  and `out', respectively. $A_{\rm s}(\tau)$ is an external source amplitude with the carrier frequency $\omega_0$.

Parameter $\Delta n_{nl}$ in Eq.~(\ref{NonCef}) is a normalization factor of the intensity dependent refractive index. It is known that in $\chi^{(3)}$ nonlinear materials it could reach a maximum value up to $10^{-4}$~\cite{Boyd:2008:NonlinearOptics}, so if we choose $\Delta n_{\rm nl}=10^{-4}$, the absolute value of $A(z,\tau)$ should not exceed unity in realistic materials.
The function $f(z)$ determines the coupling between the SNAP resonator and the microfiber. It can be approximated by a $\delta$-function~\cite{Sumetsky:2012-22537:OE},
however in order to simplify numerical simulations, here we approximate it as
%
	$f(z)={2}a^{-1} {\pi}^{-1/2}\exp[-4 (z Z - z_1)^2 a^{-2}]$,
%
where $z_1$ is the microfiber position and $a$ is the coupling width of the order of the microfiber waist.


In order to use the SNAP fiber resonator to generate frequency combs, it needs to support an equidistant spectrum. This can be achieved with a parabolic-type potential schematically illustrated in Fig.~\ref{Fig1}(b):
\begin{equation}\label{potent}
  \Delta r(z)=
  \begin{cases}
    \Delta r_{\max}\left(1 - {4z^2Z^2}{W^{-2}} \right), & |zZ|\leq W/2 \\
    0, & \mbox{otherwise}
  \end{cases}
\end{equation}
where the maximum of the potential corresponds to $\Delta r_{\max}$  radius modulation and the potential width is $W$.
The eigenfrequencies $\Omega_j$ of the potential with small mode indices $j$ can be approximated by the expression for an infinite parabolic potential as
%
	$\Omega_j \cong {\sqrt{2} c}{n_0^{-1} W^{-1}} \sqrt{{\Delta r_{\max}}/{r_0}}
            (2j-1) - \omega_0 {\Delta r_{\max}}/{r_0}$.
%
%
Then, the free spectral range is
\begin{equation}\label{FSR}
	\Delta \Omega \equiv |\Omega_j - \Omega_{j-1}| \cong 2 \sqrt{\frac{2 \Delta r_{\rm max}}{r_0}}\frac{c}{n_0 W}.
\end{equation}
This formula shows that the FSR of the system can be potentially arbitrary small. In practice, it will be limited by the fabrication capabilities. In particular, it was shown that the radius variation can be controlled with sub-angstrom precision~\cite{Sumetsky:2012-27896:OE}, and parabolic potentials with few-$nm$ maximum radius variation can be realized with high accuracy~\cite{Sumetsky:2013-163901:PRL}.

We perform numerical simulations using experimentally relevant physical parameters~\cite{Sumetsky:2013-163901:PRL}:
$r_0=19\, \mu{\rm m}$, $\lambda_0=1.5025\, \mu{\rm m}$, $\nu_0\equiv\omega_0/2\pi\approx200.5\ {\rm THz}$, loss coefficient $\gamma=0.1 pm$, $n_0=1.46$.
Parameters $C$ and $D$  were chosen as in~\cite{Sumetsky:2012-22537:OE} $C=2\times 10^4 {\rm m}^{-1}$, and $D=10^4 {\rm m}^{-1}$. The coupling width is $a =  1 \mu m$.
We analyze a mode with azimuthal index $m=116$.
%
%
We take the shape of the parabolic potential with $\Delta r_{\max}=2.8$~nm and $W=3$~mm, which was demonstrated experimentally in Ref.~\cite{Sumetsky:2013-163901:PRL}. For such a potential, the FSR is $\Delta \Omega / (2 \pi) \simeq 374 {\rm MHz}$ according to Eq.~(\ref{FSR}).
%
%
We choose the scaling parameters as $T=1 {\rm ns}$ and $Z=1\ \mu\rm{m}$, and determine the
dimensionless coefficients in \eqref{Sum3} as
$\beta_2=16816$, $\alpha=17.7$, and $\Gamma=0.083$.
We calculate the Q-factor as $\omega_0 / (\Gamma T^{-1}) = 2 \pi c T / (\Gamma \lambda_0) \sim 1.5 \times 10^{7}$.
It is also instructive to estimate the nonlinear figure of merit. Considering a normalized amplitude $|A| = 1$ and the corresponding physical nonlinear refractive index change  $\Delta n_l= 10^{-5} \div 10^{-4}$, we find that the ratio of nonlinear phase modulation and linear loss coefficients is $\alpha |A|^2 /\Gamma = 21.32 \div 213.2$, which is quite promising for observation of nonlinear effects.

\begin{figure}[tbp]
    \centering
    \includegraphics[width=\linewidth]{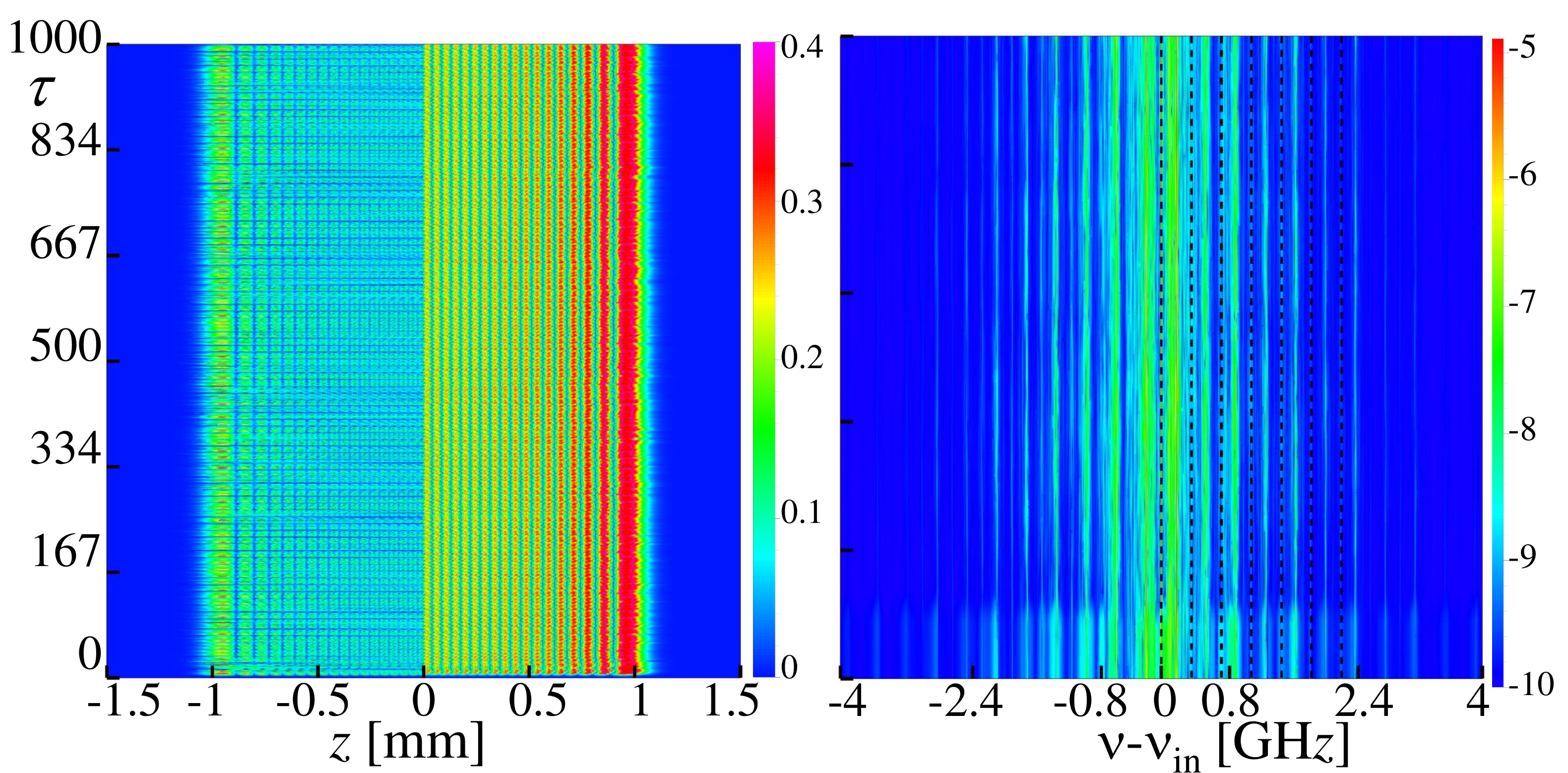}
    \caption{Generation of frequency combs in the SNAP bottle resonator from a CW input at $j=35$ eigenmode resonant frequency. (a) Intensity dynamics inside the resonator. (b) Signal spectrum evolution on the logarithmic scale, where black dashed lines indicate the parabolic potential eigenfrequencies.}
    \label{Fig2}
\end{figure}

We first present simulation results for the case of microfiber positioned in the center of the SNAP fiber parabolic potential ($z_1=0$). We consider a CW source with the frequency corresponding to the 35-th potential eigenlevel,  $A_{\rm s}(\tau)=0.8\exp({-i\Omega_{35} T \tau})$. This eigenlevel is in the middle of the equidistant spectrum, and thus satisfies the conditions for cascaded frequency comb generation.
Then, we apply a split step method~\cite{Press:2007:NumericalRecipes} to numerically solve Eqs.~(\ref{Sum3}).
%
The intensity dynamics inside the resonator is shown in Fig.~\ref{Fig2}(a). We observe periodic mode beating in the region of $|z| < 1 mm$ inside the potential well, which is a signature of comb generation.
We note that the transmitted signal is defined by the mode amplitude at the microfiber position, $A(\tau,z=z_1)$~\cite{Sumetsky:2012-22537:OE}, and calculate the frequency spectra for each time point $\tilde{\tau}$ by applying the Fourier transform to $A(\tau,z=z_1) \times w_f(\tau-\tilde{\tau})$. Here the window function is $w_f(\tau)=
1$ for $|\tau|\le b/2$, and $w_f(\tau) = \exp[-\ln16(2|\tau|/b-1)^2]$ for $|\tau| > b/2$,
and the window width was chosen at 10$\%$ of the overall time of modulation as $b=30$. We present the time evolution of the spectrum in Fig.~\ref{Fig2}(b) in the logarithmic scale with a cut off $10^{-10}$. We see that the comb is formed by approximately fifteen modes with sub-GHz frequency spacing, and it remains stable over time.

\begin{figure}[bt]
    \centering
    \includegraphics[width=\linewidth]{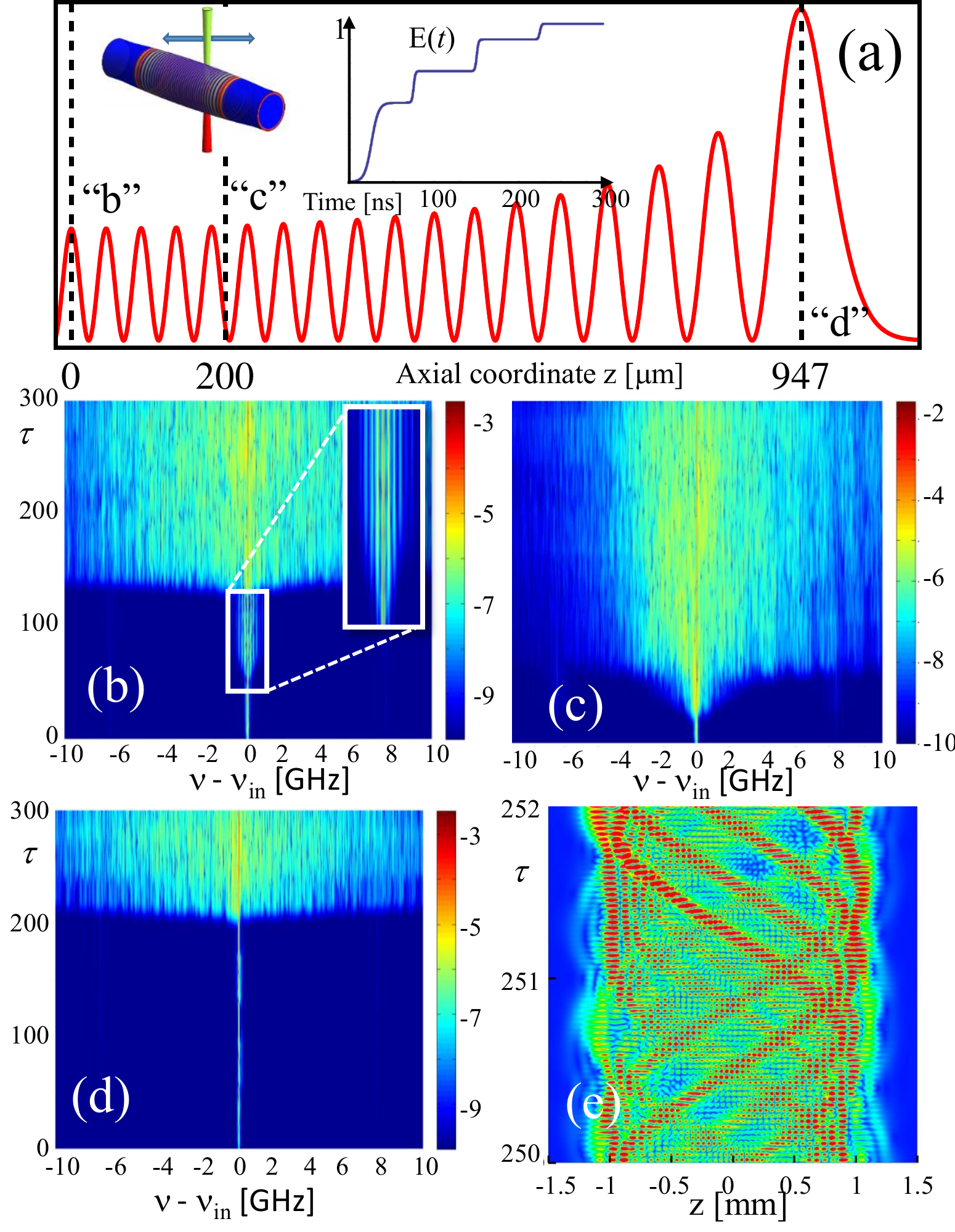}
    \caption{Nonlinear spectral broadening controlled by the microfiber position. (a) Microfiber positions (dashed lines) overlaid on the 35-th eigenmode intensity profile (red line). Inset shows the source intensity variation in time. (b-d) Time evolution of the generated spectrum for the microfiber positions indicated in (a) by dashed lines and corresponding labels. (e) Intensity dynamics in the SNAP resonator, corresponding to a noisy comb in (d) for $\tau\approx250$.}
    \label{Fig3}
\end{figure}

We find that the microfiber position can significantly affect the nonlinear dynamics.
The effect is nontrivial since the coupling between the external source and the resonator modes depends on their overlap.
We consider three representative positions: in the middle of the resonator, close to one of the mode minima, and at the global maximum of the excited mode, as indicated in Fig.~\ref{Fig3}(a). We also increase the input wave intensity during the simulation to track its effect on spectral broadening, see the inset in Fig.~\ref{Fig3}(a). The frequency spectra for the three cases are presented in the panels (b-d), respectively. For all cases we observe a spectral broadening when the source amplitude reaches some critical value. When the microfiber is placed in the center of the resonator, we observe the generation of a stable frequency comb at the time interval 50-150~ns corresponding to intermediate input intensities, as shown in Fig.~\ref{Fig3}(b), see the main figure and inset. In other two cases we observe nonstationary spectral dynamics [Figs.~\ref{Fig3}(c) and~(d)]. For the microfiber position at the global maximum of the excited mode the spectral broadening happens earlier [Fig.~\ref{Fig3}(d)], however the simulations show that nonlinearity quite rapidly deforms the "linear" mode, so that the maximum of nonlinear mode shifts and its coupling to the source ceases to be efficient. In contrast, for coupling at the center of the resonator, the even mode symmetry is preserved in the nonlinear regime, and efficient coupling is sustained at higher input intensities, facilitating stable comb generation.
Interestingly, we find that nonstationary comb dynamics is due to the excitation of moving solitons inside the resonator, see an example in Fig.~\ref{Fig3}(e).
We also observe that the number of solitons gradually increases at higher input intensities.

\begin{figure}[bt]
\centering
\includegraphics[width=\linewidth]{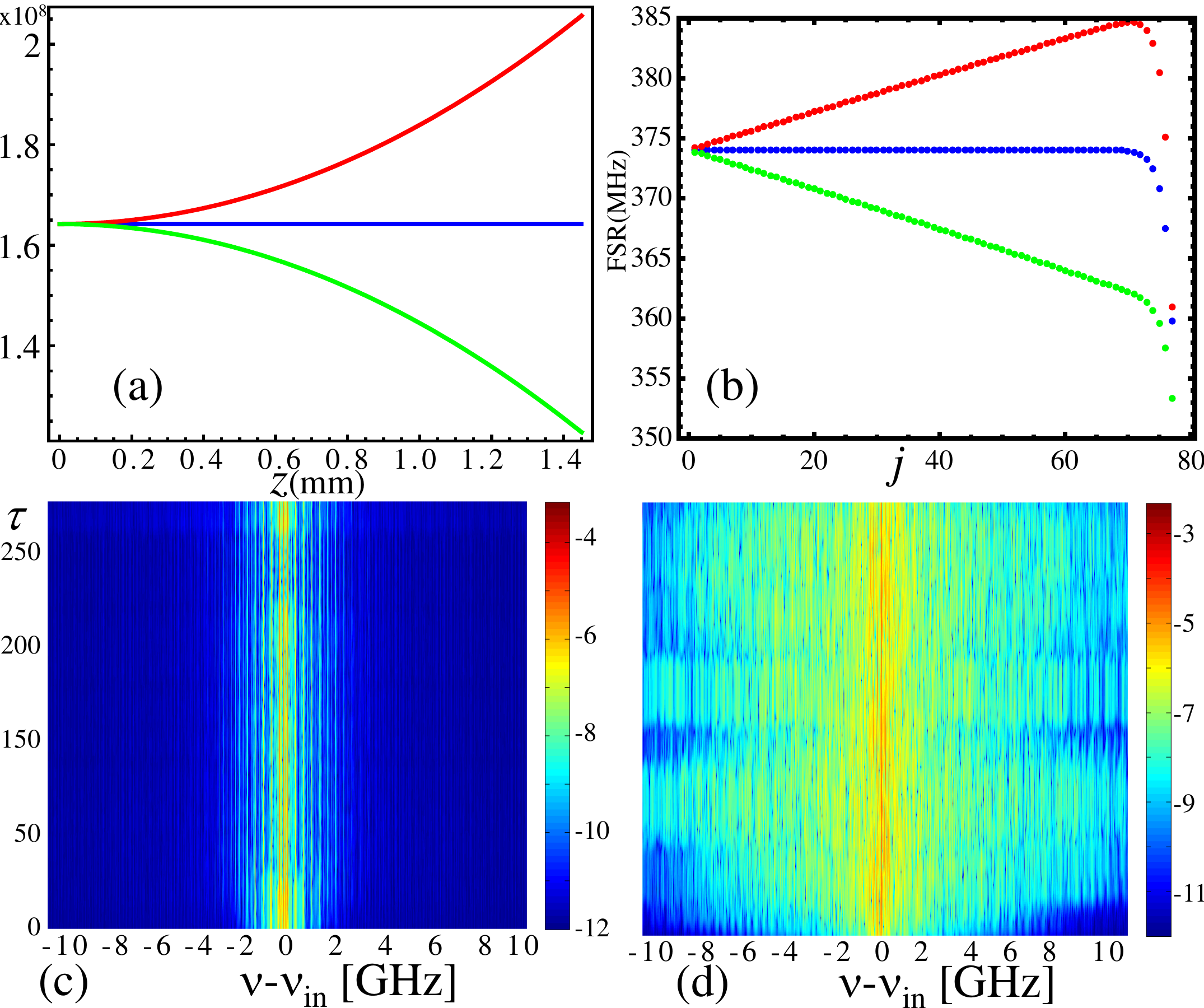}
\caption{Effect of the potential anharmonicity.
(a)~Potential curvature and (b)~free spectral range of harmonic ($B=0$, blue line) and anharmonic potentials with effectively positive FSR ($B=2\times 10^4$, red line) and negative FSR dispersion ($B=-2\times 10^4$, green line).
(c,d) Spectral evolution for the (c)~red and (d)~green anharmonic potentials. Other parameters are the same as in Fig.~\ref{Fig2}. }
\label{Fig5}
\end{figure}

We now discuss an approach to control the comb generation by shaping the potential profile. It is known that nonlinearity effectively shifts the potential eigenfrequencies and introduces additional dispersion to the system. We analyze a possibility to compensate for this effect by introducing a fourth-order correction to the parabolic potential as
\begin{equation}\label{potent4}
  \Delta r(z)=\Delta r_{\rm max}
  \begin{cases}
    1-\frac{4}{W^2}[B (z^4-z_0^4)+z^2], & |z Z|\leq \frac{W'}{2} \\
    0, & \mbox{otherwise}
  \end{cases}
\end{equation}
where $W'$ is the potential width.
Overall, such a deviation from the parabolic form causes shifts of the eigenlevels and hence changes the free spectral range.
Parameter $z_0$ gives a constant offset which does not affect the comb dynamics. For convenience, we choose $z_0 \approx 1000$, to keep approximately the same potential width around the resonant frequency of the excited eigenmode $\Omega_{35}$.
Figure~\ref{Fig5}(a) represents a curvature of the potential, $V''$. The absolute value of parameter $B$ determines the strength of anharmonicity, while its sign defines a positive or negative slope of free spectral range, as illustrated in Fig.~\ref{Fig5}(b).
On the one hand, if the potential is bent `in' ($B > 0$), the system has positive FSR dispersion. On the other hand, a negative slope (for  $B < 0$) emulates negative FSR dispersion. We note that in all three cases the slopes remain constant up to about the 70-th eigenlevel, while the input frequency is tuned to the 35-th level.

To reveal an effect of unharmonicity we study frequency comb generation in the system with  the parameters as in Fig.~\ref{Fig2}, but for the potential determined by~\eqref{potent4}. Numerical simulations demonstrate that for the positive FSR dispersion ($B=2\times10^4$) the comb spectrum does not change much [see Fig.~\ref{Fig5}(c)], while for the negative FSR dispersion ($B=-2\times10^4$) there is a significant spectral broadening~\cite{DelHaye:2007-1214:NAT}, although the comb is less stable in this case [see Fig.~\ref{Fig5}(d)]. We note that in ring resonators a positive FSR dispersion usually causes spectral broadening due to a compensation of opposite contributions from the nonlinearity-induced and material dispersions~\cite{Savchenkov:2011-293:NPHOT}. However in our case the is an additional effect of nonlinearly induced deformation of the SNAP potential, resulting in a reshaping of the resonant eigenmodes and modification of their coupling with the microfiber, which significantly affects the comb generation.


In summary, we
predict that micro-scale SNAP fiber resonators can facilitate the generation of frequency combs with ultra-fine spacing by employing the axial modes. This can be achieved by engineering a quadratic potential profile in the axial direction, and we present numerical simulations of 374~MGz comb in a resonator with a 19~$\mu m$ radius under practical conditions.
Furthermore, the SNAP potential can be arbitrarily tailored, which offers  opportunities to compensate nonlinear dispersion and control the comb dynamics.

\paragraph{Funding.}
Australian Research Council (Discovery Project DP160100619); Royal Society (WM130110); Horizon 2020 (H2020-EU.1.3.3, 691011); EPSRC (EP/P006183/1).

\paragraph{Acknowledgment.}
We thank Prof. Michael L. Gorodetsky for fruitful discussions.

\bibliography{db_FCGinSNAPfv4}


\end{document}